\documentclass[12pt]{article}
\input epsf
\newcommand{\sect}[1]{\section{#1}\setcounter{equation}{0}}

\begin{document}
\date{\today}

\def\be{\begin{equation}}
\def\ee{\end{equation}}
\def\bea{\begin{eqnarray}}
\def\eea{\end{eqnarray}}
\def\e{\epsilon}
\def\cF{{\cal F}}
\def\cN{{\cal N}}
\def\t{{\tau}}
\def\w{{\omega}}
\def\d{{\triangle}}
\def\cA{{\cal A}}

\newpage
\bigskip
\hskip 3.7in\vbox{\baselineskip12pt
\hbox{NSF-ITP-00-122}
}

\bigskip\bigskip

\centerline{\large \bf Finite temperature resolution of the Klebanov-Tseytlin
singularity}

\bigskip\bigskip

\centerline{{\bf
Alex Buchel\footnote{buchel@itp.ucsb.edu}}}

\bigskip
\centerline{Institute for Theoretical Physics}
\centerline{University of California}
\centerline{Santa Barbara, CA\ \ 93106-4030, U.S.A.}

\begin{abstract}
\baselineskip=16pt
Naked singularities in the gravitational backgrounds 
dual to gauge theories can be hidden behind the black 
hole horizon. 
We present an exact black hole solution in the Klebanov-Tseytlin 
geometry [hep-th/0002159]. Our solution 
realizes Maldacena dual of the finite temperature 
$\cN=1$ duality cascade of [hep-th/0007191] above the 
temperature of the chiral symmetry breaking. We compare 
the Bekenstein-Hawking entropy of the black hole with 
the entropy of the $SU(N+M)\times SU(N)$ gauge theory.

\bigskip
{\bf NOTE ADDED}: 
The non-extremal generalization of the KT background proposed 
in this paper was obtained partly with numerical methods.
In [hep-th/0102105] this solution has been rederived analytically, 
and  a numerical error pointed that led to the conclusion 
of a non-singular horizon in the non-BPS background discussed here.
In [hep-th/0102105] it is shown that within the self-dual ansatz 
for the three form fluxes the unique non-extremal 
solution (discussed in this paper) always has singular horizon 
for any value of non-extremality. Thus,  
the constant dilaton  ansatz (following from the self-duality of the 
three form fluxes) appears to be too restrictive to describe the 
high temperature phase of the cascading gauge theory.
The system of second order equations in the radial variable
describing non-extremal generalizations of the KT background 
whose solutions may have regular horizons is given in 
[hep-th/0102105]. (Added) section 5 of this paper discusses 
numerical error of the previous version, identified in 
[hep-th/0102105].

\end{abstract}
\newpage
\setcounter{footnote}{0}

\baselineskip=17pt

\sect{Introduction}

The AdS/CFT duality of Maldacena \cite{juan} relating supergravity  
and  strongly coupled superconformal gauge theories is usually motivated 
by comparing stacks of elementary branes with corresponding gravitational 
backgrounds in string or M-theory. For example, a large number $N$
of coincident D3-branes and the 3-brane classical solution 
leads to the duality between $\cN=4$ supersymmetric gauge 
theory and Type IIB strings on $AdS_5\times S^5$ \cite{juan,edholo,GPK}.

Subsequently, the duality has been extended to  
nonconformal systems and to systems with less supersymmetry
\cite{agmoo}. One way to break supersymmetry of the world-volume theory 
of the D3-branes is to place them at appropriate conical singularities
\cite{c1,c2,c3,c4}. 
Then the background dual to the CFT on the D3-branes is 
$AdS_5\times X_5$ where $X_5$ is the Einstein manifold which is the 
base of the cone. Breaking conformal invariance without 
further breaking the supersymmetry  can be achieved by 
placing fractional D3 branes at the singularity in addition 
to the regular ones. These fractional D3 branes are D5-branes 
wrapped over (collapsed) 2-cycle at the singularity \cite{jT}. 
In the case of a conifold the singularity is a point. 
Placing $M$ fractional and $N$ regular D3-branes Klebanov and Nekrasov 
constructed \cite{KN} the renormalization group flow in the gravity 
dual to the $\cN=1$ supersymmetric $SU(N)\times SU(N+M)$ gauge 
theory\footnote{Renormalization group flows on fractional D3-branes 
at an orbifolded conifold were discussed in \cite{ot}.}. The supergravity equations were solved to leading order 
in $M/N$ in \cite{KN}. In \cite{KT},  this solution (which we refer 
to from now on as KT) was completed to 
all orders; the conifold suffers logarithmic warping, and 
the relative gauge coupling of the two gauge factors 
runs logarithmically at all scales. The D3-brane charge, i.e 
the 5-form flux, decreases logarithmically as well. However, 
the logarithm in the solution is not cut off at small radius; 
eventually the D3-charge becomes negative and one encounters 
a naked singularity  in the metric. 

The appearance of naked singularities in the gravitational 
dual of the nonconformal gauge theories with reduced supersymmetry 
is rather common phenomenon. Thus, understanding the physics of these 
singularities is an important problem. By now 
we know several different mechanisms for resolving singularities,
which seem to depend on the amount of supersymmetry present in 
the problem. In the case when the gauge theory has 8 supercharges 
($\cN=2$ supersymmetry in $D=4$) the naked singularity, 
known as repulson, is resolved by the enhancon mechanism 
characterized by the expansion of a system of branes in the 
transverse directions \cite{JPP,JJ,PW,BPP,EJP,bvflmp}. In the 
class of $\cN=1$ gauge theories obtained by mass deformation 
of parent $\cN=4$ gauge theories, the naked singularity of the 
gravitational dual is resolved by Dp-brane polarization 
into a spherical Dp+2-brane \cite{PS,B,AR,BN1,BN2,clp} 
via Myer's dielectric effect \cite{M}. A common 
theme in the enhancon and the brane polarization
mechanism is that the singularity of the gravitation 
background is resolved by a distinctive string/braney 
phenomena. The resolution of the singularity in the 
KT geometry is rather different. Klebanov and Strassler \cite{ks0007}
showed that in this case the resolution is achieved 
entirely within supergravity by deforming the conifold. 
The crucial observation made in \cite{ks0007} 
was the identification of the 5-form flux decrease  in KT with 
the repeated chain of Seiberg-duality transformations 
in which $N\to N-M$. At the IR end of the duality cascade the 
gauge theory confines with the chiral symmetry breaking. 
Klebanov and Strassler convincingly argued that the 
chiral symmetry breaking on the gauge theory side is 
dual to deforming the conifold on the gravity side of the Maldacena 
duality. 

The resolutions of singularities we mentioned above, are for the 
gravitational backgrounds dual to the gauge theories at zero 
temperature. There is an extensive literature\footnote{See \cite{agmoo}
for discussion and references.} on the application of gravity/gauge 
theory duality to the finite 
temperature gauge theories starting with the work of Witten 
\cite{w9803}. Finite temperature gauge theories have 
black hole gravitational dual. Thus, one can imagine 
that by ``heating'' the gauge theory, the naked singularity 
of its gravitational dual will be hidden behind the black hole 
horizon. 
In fact, this underlines the classification of naked 
singularities proposed in \cite{g0002}\footnote{In 
\cite{g0002} this criterion
has been applied for admissibility of 
singular solutions with four dimensional Poincare 
invariance in five-dimensional gravity with scalars.}: the ``good'' naked 
singularities  are those which can be obtained as limits of 
regular black holes. 
One of the motivations for the criterion \cite{g0002} 
is that finite temperature in field 
theory serves as an infrared cutoff in the sense that it masks physics 
at scales lower than the temperature. A naked singularity that can 
be hidden behind a black hole horizon is a signal of a non-trivial 
but sensible infrared physics in the dual gauge theory. In particular, 
this might signal of a phase transition in the gauge theory.
If we approach a ``good'' naked singularity by taking a 
limit of regular black holes, the dual picture is that the infrared 
cutoff (finite temperature) is being removed. 
As the temperature is sufficiently lowered, the horizon 
should retreat until the original singular geometry is recovered.   
For the  KT singularity we can turn the 
argument around: since we {\it know} that KT singularity is 
``good''  --- we know of its resolution --- there {\it must}
exist a black hole solution with the asymptotic KT geometry.
Another way to reach the same conclusion is to recall 
that at zero temperature, the naked singularity of the 
gravitational dual was resolved by the mechanism dual to the 
chiral symmetry breaking in the gauge theory. However,
at high enough temperature, we expect restoration 
of the chiral symmetry in the gauge theory. Thus, the Maldacena 
dual of this gauge  theory at temperature 
above chiral symmetry breaking should be described by a
black hole in  KT geometry\footnote{Note that the chiral symmetry breaking 
phase transition is precisely the example of a non-trivial 
infrared dynamics which can be hidden by the  IR cutoff.}. 
In this paper we construct such a black hole solution. 

A finite temperature generalization of  Polchinski-Strassler
construction  \cite{PS} 
has been recently discussed in 
\cite{fm0007}. It was also emphasized there that
in the high temperature phase it is unnecessary to invoke 
Myers' mechanism since there are no naked singularities. 
The gravity solution was found in \cite{fm0007} in the leading 
order in mass perturbation.  

This paper is organized as follows. We start the next section 
by introducing our conventions and recalling KT solution. 
We then describe an exact black hole solution in KT geometry.
In section 3 we compute the Bekenstein-Hawking entropy of this black 
hole and compare it to the entropy of the dual gauge theory. 
We conclude in section 4.

\section{Black hole solution in KT geometry}

\subsection{Type IIB equations of motion}

We use mostly negative conventions for the signature 
$(+-\cdots -)$ and $\epsilon^{1\cdots10}=+1$.
The type IIB equations consist of \cite{schwarz83}:

\noindent $\bullet$\quad  The Einstein equations:
\be
R_{MN}=T^{(1)}_{MN}+T^{(3)}_{MN}+T^{(5)}_{MN}\,,
\label{tenein}
\ee
where the energy
momentum tensors of the dilaton/axion field, $B$, the three index
antisymmetric tensor field, $F_{(3)}$, and the self-dual five-index
tensor field, $F_{(5)}$, are given by
\be
T^{(1)}_{MN}= P_MP_N{}^*+P_NP_M{}^*\,,
\label{enmomP}
\ee
\be
T^{(3)}_{MN}=
       {1\over 8}(G^{PQ}{}_MG^*_{PQN}+G^{*PQ}{}_MG_{PQN}-
        {1\over 6}g_{MN} G^{PQR}G^*_{PQR})\,,
\label{enmomG}
\ee
\be
T^{(5)}_{MN}= {1\over 6} F^{PQRS}{}_MF_{PQRSN}\,.
\label{enmomF}
\ee

In the unitary gauge $B$ is a complex scalar
field and
\be
P_M= f^2\partial_M B\,,\qquad Q_M= f^2\,{\rm Im}\,(
B\partial_MB^*)\,,
\label{defofPQ}
\ee
with
\be
f= {1\over (1-BB^*)^{1/2}}\,,
\label{defoff}
\ee
while the antisymmetric tensor field $G_{(3)}$ is given by
\be
G_{(3)}= f(F_{(3)}-BF_{(3)}^*)\,.
\label{defofG}
\ee

\noindent 
$\bullet$\quad The Maxwell equations:
\be
(\nabla^P-i Q^P) G_{MNP}= P^P G^*_{MNP}-{2\over 3}\,i\,F_{MNPQR}
G^{PQR}\,.
\label{tenmaxwell}
\ee

\noindent
$\bullet$\quad The dilaton equation:
\be
(\nabla^M -2 i Q^M) P_M= -{1\over 24} G^{PQR}G_{PQR}\,.
\label{tengsq}
\ee
\noindent
$\bullet$\quad The self-dual equation:
\be
F_{(5)}= \star F_{(5)}\,.
\label{tenself}
\ee

In addition, $F_{(3)}$ and $F_{(5)}$ satisfy Bianchi identities which
follow from the definition of those field strengths in terms of their
potentials:
\bea
F_{(3)}&=& dA_{(2)}\,,\cr
F_{(5)}&=& dA_{(4)}-{1\over 8}\,{\rm Im}( A_{(2)}\wedge
F_{(3)}^*)\,.
\label{defpotth}
\eea

We consider the background with constant dilaton and zero axion 
\be
B={1-g_s\over 1+g_s}\,.
\label{bgs}
\ee

We also use
\be
G_3=f (1-B) \cF_3+i f (1+B)  H_3\,,
\label{g3fh}
\ee
where $\cF_3$ and $H_3$ are the RR and NSNS 3-form field strength 
correspondingly.

\subsection{KT solution}

In this section we review KT \cite{KT} gravitational dual of the 
$SU(N)\times SU(N+M)$ gauge theory, mainly to set up 
our notations, and discuss the dictionary between two descriptions. 

The Einstein frame metric is:
\be
ds^2_{10}=c_1(\t)^2\left[dt^2-d\bar{x}^2\right]
-c_2(\t)^2\left[d\t^2+9\ ds^2_{T^{1,1}}\right]\,,
\label{m10kt}
\ee
where  $ds^2_{T^{1,1}}$ is the metric on the coset space 
$T^{1,1}=(SU(2)\times SU(2))/U(1)$ which is the base of a cone 
\cite{CO}  
\be
ds^2_{T^{1,1}}={1\over 9} (g^5)^2+{1\over 6}\sum_{i=1}^4 (g^i)^2\,.
\label{t11}
\ee
We use the 
following basis of 1-forms on the compact space \cite{MT}
\bea
g^1&=&\left(-\sin\theta_1 d\phi_1-\cos\psi\sin\theta_2 d\phi_2+\sin\psi
d\theta_2\right)/\sqrt{2}\,,\cr
g^2&=&\left(d\theta_1-\sin\psi\sin\theta_2 d\phi_2-\cos\psi d\theta_2
\right)/\sqrt{2}\,,\cr
g^3&=&\left(-\sin\theta_1 d\phi_1+\cos\psi\sin\theta_2 d\phi_2-\sin\psi
d\theta_2\right)/\sqrt{2}\,,\cr
g^4&=&\left(d\theta_1+\sin\psi\sin\theta_2 d\phi_2+\cos\psi d\theta_2
\right)/\sqrt{2}\,,\cr
g^5&=&d\psi+\cos\theta_1 d\phi_1+\cos\theta_2 d\phi_2\,.
\label{forms}
\eea
Furthermore, 
\bea
c_1(\t)&=&H(\t)^{-1/4}\,,\cr
c_2(\t)&=&\e\ H(\t)^{1/4}\ {e^{\t/3}\over 3}\,,\cr
\cr
H(\t)&=&b_0+{81\over 2} g_s M^2 e^{-4\t/3}\left[{\t\over 3}-\ln{r_s\over\e}
\right]\,.
\label{warpKT}
\eea
Introducing a new radial variable 
\be
r=\e\ e^{\t/3}\,,
\label{rtau}
\ee
and eliminating asymptotically flat region (setting $b_0=0$) we can 
rewrite (\ref{m10kt}) as
\be
ds^2_{10}=H(r)^{-1/2} \left[dt^2-d\bar{x}^2\right]-H(r)^{1/2}
\left[dr^2+ r^2\ ds^2_{T^{1,1}}\right]\,,
\label{rKT}
\ee
with 
\be
H(r)={L^4\over r^4}\ln{r\over r_s}\,,\qquad L^4={81\over 2}g_s M^2\e^4\,.
\label{rh}
\ee
As we already mentioned, there is a naked singularity in KT geometry.
From (\ref{rh}) we see that it is at $r=r_s$.

The dilaton is constant, and the three-form field strength
\be
G_3={1\over 2}g_s^{1/2} M\biggl[g_3\wedge g_4\wedge g_5+
g_1\wedge g_2\wedge g_5 - i d(\t)\wedge g_3\wedge g_4 -id(\t)\wedge g_1\wedge 
g_2   \biggr]\,.
\label{g3}
\ee
is self-dual
\be
G_3=i\star_6 G_3\,.
\label{duality}
\ee
Finally, the five-form field strength is 
\be
F_5=\cF_5+\star \cF_5\,,
\label{5form}
\ee
with 
\be
\cF_5={1\over 4} \left[c_1(\t)^4\right]'\ dt\wedge dx_1\wedge dx_2\wedge dx_3
\wedge d(\t)\,.
\label{cf5}
\ee
In (\ref{cf5}) (and below) the prime denotes the derivative with 
respect to $\t$.

The dictionary between gauge and gravity descriptions \cite{KN,KT,ks0007} 
is most clear  when one uses $r$ (\ref{rtau}) as a radial 
coordinate on the gravity side; $\mu\equiv r/\e^2$ is then natural to identify 
with the RG scale of the gauge theory. On the gauge theory 
side, at a given energy scale $\mu$, we have $SU(N+M)\times SU(N)$ 
gauge theory with two chiral superfields $A_1,A_2$ in the $(N+M,\overline{N})$
representation, and two fields $B_1,B_2$ in the $(\overline{N+M},N)$ 
representation. The superpotential of the model is 
\be
W\sim {\rm tr}\ (A_iB_jA_kB_\ell)\ \e^{ik} \e^{j\ell}\,.
\label{superpotential}
\ee
The $SU(2)\times SU(2)\times U(1)$ global symmetry of the 
$T^{1,1}$ is realized as a global symmetry of the gauge theory
with the first (second) $SU(2)$ factor rotating the flavor 
index of the $A_i$ $(B_i)$, while the ``baryon'' $U(1)$
acts as $A_i\to A_i e^{i\alpha}$, $B_i\to B_i e^{-i\alpha}$. 
The two gauge group factors have gauge  couplings $g_1$
and $g_2$. Under the RG flow, the sum of the couplings
does not run. It is dual to the 
dilaton of in  KT background:
\be
{1\over g_s}={4\pi\over g_1^2}+{4\pi\over g_2^2}\,.
\label{sum}
\ee 
On the gauge theory side, the difference between two couplings 
is 
\be
{4\pi\over g_2^2}-{4\pi\over g_1^2}\sim M\ln(\mu/\Lambda) [3+2(1-\gamma)]\,,
\label{difference}
\ee
where $\gamma$ is the anomalous dimension of operators ${\rm tr}\ A_iB_j$.
This is dual on the gravity side to the NSNS 2-form flux $B_2$ through the 
$S^2$ of the cone base:\footnote{Topologically $T^{1,1}$ is $S^2\times
S^3$.}
\be 
{1\over g_s}\ \int_{S^2} B_2 \sim  M \ln(r/\e) \sim   M \ln(\mu/\Lambda)\,,
\label{b2flux}
\ee
where we identified $\Lambda\sim 1/\e$.
It  is clear from (\ref{sum}), (\ref{difference}) that starting 
at some energy scale and flowing either to the UV or the IR   
one inevitably hits a Landau pole --- one of the two gauge couplings 
will become infinitely large. Klebanov and Strassler argued 
that this singularity in the gauge theory is artificial, 
and arises as one insists on describing strongly coupled 
gauge dynamics through the perturbative degrees of freedom.
Rather, \cite{ks0007} one has to do a Seiberg duality \cite{s9402}
on the strongly coupled gauge factor, which becomes weakly 
coupled in terms of a dual, ``magnetic'' description.   
We will not discuss the details here; only collect 
the necessary facts. Turns out that the Seiberg duality 
transformation is a self-similarity transformation of the 
gauge theory which replaces $N\to N-M$ as one flows to the IR,
or $N\to N+M$ as the energy increases. Thus, effectively,
the rank of the gauge group is not constant along the RG 
flow, but changes with energy:
\be
N=N(\mu)\sim M^2\ \ln(\mu/\Lambda )\,.
\label{nmu}
\ee
(To see (\ref{nmu}) note that the rank changes by $\triangle N\sim M$
as $M\triangle (\ln(\mu/\Lambda))\sim 1$.) This fits nicely 
with the result of  the gravitational dual where  the five-form 
field strength with components along  the $T^{1,1}$ grows as 
\be
F_5=\cF_5+\star\cF_5\,,\qquad \star\cF_5\sim g_s M^2\ \ln(r/\e) 
{\rm vol}(T^{1,1})\,.
\label{55form}
\ee 
The original Seiberg duality \cite{s9402} was proposed between 
two different gauge theories in the UV which flow in the 
IR to the same superconformal fixed point. The duality 
cascade of Klebanov-Strassler is an extension of the 
electric-magnetic Seiberg duality in several aspects\footnote{
I would like to thank Matt Strassler for discussing 
this point.}. First of all,  KS duality is an equivalence between 
nonconformal theories. Here, there is no anomaly-free $U(1)_R$ 
symmetry which in the case of Seiberg duality was essential 
to match chiral rings of two theories. From the purely 
field theoretic perspective, one also can not 
use chiral anomalies of global symmetries to argue for the equivalence 
of two dual descriptions --- the only unbroken global 
symmetries of the problem are $SU(2)\times SU(2)\times U(1)$ symmetry 
of the conifold: $SU(2)$ symmetries don't have chiral anomalies 
and the ``baryon'' $U(1)$ is not chiral. Second, 
unlike the original Seiberg duality which is true in the far IR, 
dualities of KS arise at finite energies. Third, in \cite{ks0007}
duality has been performed on only one gauge factor while the other 
one was a spectator. Strictly speaking, to argue for the duality 
the coupling of that gauge factor has to be exactly zero. However, 
 KS duality changes couplings of both gauge factors, thus the 
latter assumption seems inconsistent. 

Above arguments indicate that conventional tools used to check
dualities of $\cN=1$ gauge theories are not very useful for KS 
dualities. The strongest support for KS duality cascade 
comes from the behavior of the five-form filed strength of 
its gravitational dual. In section 3 of this paper we present 
another evidence in favor of KS duality cascade:
we show that the Bekenstein-Hawking entropy of a black 
hole in  KT geometry matches (up to numerical coefficients) 
the entropy of the $SU(N(\mu))\times SU(N(\mu)+M)$ gauge theory.

\subsection{Black hole solution in  KT geometry}
Black holes have long been objects of interests in string 
theory\footnote{For an introduction to the subject and references 
see \cite{peet}.}. 
In the context of AdS/CFT  duality, they realize gravitational
dual of gauge theories at finite temperature. In this section 
we present an exact  solution of type IIB equations of motion 
describing a black hole in KT geometry. The solution 
realizes the gravitational dual of  KS cascade of 
dualities \cite{ks0007} at temperatures above chiral symmetry breaking;
it is also the solution which can ``hide'' naked singularity 
of  KT geometry.

We start with the following ansatz for the metric
\be
ds^2_{10}=c_1(\t)^2\left[\d_1(\t)^2  dt^2-d\bar{x}^2\right]
-c_2(\t)^2\left[\d_2(\t)^{-2} d\t^2+9\ ds^2_{T^{1,1}}\right]\,,
\label{ktt}
\ee
where we introduce two new warp factor $\d_1(\t)$ and $\d_2(\t)$, 
such that $\d_i(\t)\to 1$ as $\tau\to \infty$. Also we 
expect to find zeros of the warp factors at finite $\t$s.  

We will look for the solution with self-dual three form 
field strength (\ref{duality}); this  insures that the dilaton 
is constant. The RR 3-form ansatz we take is 
\bea
\cF_3&=&M g_5\wedge g_3 \wedge g_4+d\biggl[
f_1(\tau) g_1\wedge g_2+f_2(\tau) g_1\wedge g_3+f_3(\tau) g_1\wedge g_4\cr
&&+f_4(\tau) g_1\wedge g_5+f_5(\tau) g_2\wedge g_3+f_6(\tau) g_2\wedge g_4+f_7(\tau) 
g_2\wedge g_5\cr
&&+f_8(\tau) g_3\wedge g_4+f_9(\tau) g_3\wedge g_5+f_{10}(\tau) g_4\wedge g_5
\biggr]\,,
\label{F3}
\eea 
while the one for the NSNS 3-form
\bea
H_3&=&d\biggl[
h_1(\tau) g_1\wedge g_2+h_2(\tau) g_1\wedge g_3+h_3(\tau) g_1\wedge g_4\cr
&&+h_4(\tau) g_1\wedge g_5+h_5(\tau) g_2\wedge g_3+h_6(\tau) g_2\wedge g_4+h_7(\tau) 
g_2\wedge g_5\cr
&&+h_8(\tau) g_3\wedge g_4+h_9(\tau) g_3\wedge g_5+h_{10}(\tau) g_4\wedge g_5
\biggr]\,.
\label{H3}
\eea
In (\ref{F3})  $M$ is  a number. $g_5\wedge g_3 \wedge g_4$
is a  closed 3-form on $T^{1,1}$ which is not exact. 
Clearly, 
\be
dG_3=0\,.
\label{G3}
\ee
Furthermore, we use the black p-brane ansatz for the 
5-form:
\bea
F_5&=&\cF_5+\star\cF_5\,,\cr
\cr
\cF_5&=&{\w(\t) c_1(\t)^4 c_2(\t) \d_1(\t)\over \d_2(\t)}\
dt\wedge dx_1\wedge dx_2\wedge dx_3 \wedge d(\t)\,.
\label{cft}
\eea
 
With our ansatz for the 3-forms, the self-duality (\ref{duality}) 
constraints 
\bea
h_5(\tau)&=&-h_3(\tau)\,,\cr
f_5(\tau)&=&-f_3(\tau)\,,\cr
f_6(\tau)&=&f_2(\tau)\,,\cr
h_6(\tau)&=&h_2(\tau)\,,\cr
f_4(\tau)&=&0\,,\cr
h_4(\tau)&=&0\,,\cr
f_9(\tau)&=&0\,,\cr
h_9(\tau)&=&0\,,\cr
f_{10}(\tau)&=&f_7(\tau)=0\,,\cr
h_{10}(\tau)&=&h_7(\tau)=0\,,\cr
\cr
f_3'(\tau)&=&0\,,\cr
h_3'(\tau)&=&0\,,\cr
f_8'(\tau)&=&{h_2(\tau)\over g_s \d_2(\tau)}\,,\cr
h_8'(\tau)&=&-{g_s f_2(\tau)\over \d_2(\tau)}\,,\cr
f_2'(\tau)&=&{1\over 2 g_s \d_2(\tau)}\ (h_1(\tau)-h_8(\tau))\,,\cr
h_2'(\tau)&=&{g_s\over 2 \d_2(\tau)}\ (f_8(\tau)-f_1(\tau))\,,\cr
f_1'(\tau)&=&-{h_2(\tau)\over g_s \d_2(\tau)}\,,\cr\cr
h_1'(\tau)&=& {g_s\over \d_2(\tau)}\ (f_2(\tau)-M)\,.
\label{hconst}
\eea

In KT solution (from (\ref{g3}))
\bea
\d_1(\tau)&=&\d_2(\tau)=1\,,\cr
f_1(\t)&=&f_8(\t)=h_2(\t)=0\,,\cr
f_2(\t)&=&{1\over 2} M\,,\cr
h_1(\t)&=&h_8(\t)=-{1\over 2} g_s M \t\,.
\label{ks3fluxes}
\eea
We would like to have asymptotically as $\t\to \infty$ KT solution,
so we choose the following solution of (\ref{hconst})
\bea
f_1(\t)&=&f_8(\t)=h_2(\t)=0\,,\cr
f_2(\t)&=&{1\over 2} M\,,\cr
h_8(\t)&=&h_1(\t)\,.
\label{gmy}
\eea
All constraints of (\ref{hconst}) are reduced to a single ODE
\be
h_1'(\tau)=-{g_s M\over 2 \d_2(\t)}\,.
\label{h1eq}
\ee

The 3-form Maxwell equations are satisfied provided 
\be
\w(\t)={1\over 4}\ {\d_2(\t)\over c_1(\t)^4 c_2(\t) \d_1(\t)}\ 
\left[ \d_1(\t) c_1(\t)^4\right]'\,.
\label{maxwell}
\ee

Consider now the Einstein equations (\ref{tenein}).
The dilaton/axion is constant in our background, so 
\be
R_{MN}=T^{(3)}_{MN}+T^{(5)}_{MN}\,.
\label{ein}
\ee
The energy-momentum tensor of the five-form is 
diagonal traceless
\be
T^{(5)}_{MN}=4 \w(\t)^2\ diag\{1,-1,-1,-1,-1,1,1,1,1,1\}\,.
\label{t5}
\ee 
The energy-momentum tensor of the three-form is diagonal as well
\be
T^{(3)}_{MN}={g_s M^2\over 18 c_2(\t)^6}\ diag\{1,-1,-1,-1,1,1,1,1,1,1\}\,,
\label{t3}
\ee 
where we used constraints (\ref{gmy}) and  (\ref{h1eq}).
From (\ref{ein}) it now follows that 
\be
R_{11}+R_{22}=0\,, 
\label{ewarp}
\ee
which gives a differential equation relating warp factors $\d_i(\t)$:
\bea
R_{11}+R_{22}=&&0\cr
=&&{\d_2(\t)\over c_1(\t) c_2(\t)^3 \d_1(\t) }
\biggl[4 \d_2(\t) \left[\d_1(\t)\right]' \left[c_1(\t) c_2(\t)\right]' \cr
&&+c_1(\t) c_2(\t) \left[\d_2(\t)\left[\d_1(\t)\right]'\right]'\biggr]\,.
\label{d1d2}
\eea
The latter equation is easy to integrate once:
\be
\d_2(\t)\left[\d_1(\t)\right]'={A\over c_1(\t)^4 c_2(\t)^4 }\equiv h(\t)\,,
\label{drel}
\ee
where $A$ is an integration constant. For later convenience 
we introduce $h(\t)$ as in (\ref{drel}). Note that for the 
$c_i(\t)$ in the original KT solution  
\be
h(\t)={81 A\over \e^4} e^{-4\tau/3}\,,
\label{h}
\ee
and $h(\t)\to 0$ as $\t\to \infty$.
This fact will be important later, as it will turn out that 
we can use the same $h(\t)$ here.

At this stage it is convenient to identify constraint from the 
five-form Bianchi identity. Using (\ref{drel}), we find:
\bea
\left[c_1(\t)\right]''=&&{1\over 36 \d_1(\t)^2\d_2(\t)^2 c_1(\t)^7 
c_2(\t)^8} \biggl(\cr
&& 9 A^2+2 g_s M^2 \d_1(\t)^2 c_1(\t)^8  c_2(\t)^4\cr
&&+36 \d_1(\t)\d_2(\t) A c_1(\t)^3 c_2(\t)^4 \left[c_1(\t)\right]'\cr
&&+36 \d_1(\t)^2\d_2(\t)^2  c_1(\t)^6 c_2(\t)^8 
\left(\left[c_1(\t)\right]'\right)^2\cr
&&-144 \d_1(\t)^2\d_2(\t)^2 c_1(\t)^7 c_2(\t)^7 \left[c_1(\t)\right]' 
\left[c_2(\t)\right]'\cr
&&-36 \d_1(\t)^2\d_2(\t) c_1(\t)^7 c_2(\t)^8 \left[\d_2(\t)\right]'
\left[c_1(\t)\right]'
\biggr)\,.
\label{d2c1}
\eea

Now we are left with three independent 
Einstein equations (\ref{ein}) for the components $M=N=\{1,5,6\}$.  
Turns out, that given (\ref{maxwell}), (\ref{drel}) and (\ref{d2c1}), the 
$M=N=1$ Einstein equation is satisfied automatically, while 
the other two give a nonlinear system of  ordinary differential 
equations on $h(\t)$ and $\d_1(\t)$:
\bea
0=&&2h(\t)\left(\left[\d_1(\t)\right]'\right)^3-5
\left(\d_1(\t)\right)^2\left[h(\t)\right]'
\left[\d_1(\t)\right]''\cr
&&-5\d_1(\t)\left[h(\t)\right]'
\left(\left[\d_1(\t)\right]'\right)^2+5\d_1(\t)^2 \left[\d_1(\t)\right]'
\left[h(\t)\right]''\,,\cr
\cr
0=&&9 \d_1(\t)\left[\d_1(\t)\right]'\left(\left[h(\t)\right]'\right)^2
+9\d_1(\t) h(\t) \left[h(\t)\right]'\left[\d_1(\t)\right]''\cr
&&-9h(\t)\left[h(\t)\right]'\left(\left[\d_1(\t)\right]'\right)^2
-9\d_1(\t) h(\t) \left[\d_1(\t)\right]'\left[h(\t)\right]''\cr
&&-16\d_1(\t) \left(\left[\d_1(\t)\right]'\right)^3\,,
\label{e5e6}
\eea
for  $M=N=5$ and $M=N=6$ correspondingly.
A trick to solve (\ref{e5e6}) is to notice that $\d_1(\t)$ has 
only implicit $\t$ dependence through $h(\t)$, that is,
\be
\d_1(\t)=\d_1(h(\t))\,.
\label{dim}
\ee
This should not come as a surprise since from (\ref{drel})
it is clear that by changing  a radial coordinate $\tau$
, i.e. redefining $\tau\to \tilde{\tau}=\tilde{\tau}(\t)$,
one changes $h(\t)\to\tilde{h}(\t)=h(\t) d\t/d\tilde{\t}$.
One obvious solution (which eventually gives original KT solution)
is then $h(\t)=0$. As we are interested in the black hole 
solutions which have nontrivial warp factors,
we assume $\left[h(\t)\right]'\ne 0$. Eqs.~(\ref{e5e6}) 
are then reduced to two ordinary differential 
equations on a single function
\be
f(h)\equiv \left(\d_1(h(\t))\right)^2\,,
\label{f}
\ee
namely,
\bea
0=&&x \left(\left[f(x)\right]'\right)^3-10 f(x)^2 \left[f(x)\right]''\,,\cr
\cr
0=&&9 f(x)\left[f(x)\right]'-9 x\left(\left[f(x)\right]'\right)^2\cr
&&+9 x f(x) \left[f(x)\right]''-4\left(\left[f(x)\right]'\right)^3\,,
\label{eqf}
\eea
where the prime denotes the derivative with respect to $x$.
Turns out, both equations are solved simultaneously 
provided 
\be
\left[f(x)\right]'={3\over 2}\ 
{\left(30x\pm 2\sqrt{135 x^2+400 f(x)}\right) f(x)
\over 9x^2-40 f(x)}\,.
\label{fp}
\ee
First thing to note is that as we've been able to solve (\ref{e5e6})
for arbitrary $h(\t)$, we might as well use $h(\t)$ as given 
in (\ref{h}). Thus, we have a boundary condition for (\ref{fp})\footnote
{Recall that the geometry should approach that of  KT for large 
$\t$.}
\be
f(0)=1\,.
\label{bc}
\ee 
Second, the solution we need has a plus sign 
in (\ref{fp}): with the boundary condition (\ref{bc}), 
only this solution has a zero, which we find numerically to be at 
\be
f(x_\star)=0\qquad {\rm as}\qquad x_\star\approx 0.55647\,.
\label{zerof}
\ee
Third, lacking the closed analytical solution for (\ref{fp}),
we can solve it only as a power series for small $x$. For completeness, 
we present the 
first few terms in large $\t$ (equivalent to the small $x$) expansion of the 
warp factors $\d_i(h(\t))$:
\bea
\d_1(x)=&&1-{3\over 4} x -{9\over 32}x^2+O(x^3)\,,\cr
\cr
\d_2(x)=&&1-{3\over 4} x -{63\over 160}x^2+O(x^3)\,.
\label{dseries}
\eea

Given the solution to the warp factors $\d_i(\t)$, we can go back 
and compute $c_i(\t)$ (from (\ref{d2c1}) and (\ref{drel})), 
the $\w(\t)$ (from (\ref{maxwell})), 
and solve (\ref{h1eq}) for the three-form. We will not attempt 
to solve these equations exactly, but rather present the leading 
behavior of the metric at large $\t$. To make contact with KT solution, 
we use a radial coordinate as in (\ref{rtau}) and 
\be
L^4={81\over 2}g_s M^2\e^4\,,\qquad r_0^4={243 A\over 2}\,.
\label{lr0}
\ee
The metric is then given by
\bea
ds^2_{10}=H(r)^{-1/2} \left[\d_1(r)^2 dt^2-d\bar{x}^2\right]-H(r)^{1/2}
\left[{dr^2\over \d_2(r)^2} + r^2\ ds^2_{T^{1,1}}\right]
\label{final}
\eea
with 
\bea
H(r)={L^4\over r^4} \ln{r\over r_s}-{L^4 r_0^4\over 8 r^8}+\cdots\,,\cr
\d_1(r)=1-{r_0^4\over 2 r^4}+\cdots\,,\cr
\d_2(r)=1-{r_0^4\over 2 r^4}+\cdots\,,
\label{mfun}
\eea
where the dots indicate subdominant $1/r$ and $\ln(r)/r$ corrections
as $r\to \infty$. The event horizon  $r_{\star}$ 
of the stationary black hole geometry 
occurs where $g^{rr}=0\Leftrightarrow \d_2(r_\star)=0$. 
Numerically, we find 
\be
\left({r_0\over r_\star}\right)^4\equiv \xi\approx 1.02427\,.
\label{rorstar}
\ee
From (\ref{mfun}), 
the black hole horizon will ``cloak'' a naked singularity in 
KT geometry if $H(r_\star)>0$. Assuming $L\gg r_0$, this 
condition translates into 
\be
r_0> r_s\ \xi^{1/4} e^{\xi/8}\approx 1.14343\ r_s
\label{rrs}
\ee  
The fact that a black hole in KT geometry must be larger
then of certain critical size (given by (\ref{rrs})) 
to hide the naked singularity, is the gravitational dual reflection 
that chiral symmetry is restored in
the gauge theory at {\it finite} temperature. 
We return to this issue in the following section.

\section{Thermodynamics of KT geometry}

In this section we compute the entropy of the black hole discussed 
in the previous section 
and compare it with the entropy of the dual gauge theory. 

To compute the Hawking temperature of the black hole (\ref{ktt}),
we introduce the proper distance near the horizon:
\be
d\eta\approx c_2(\t_\star) \d_2(\t)^{-1} d\t\,,
\label{nu}
\ee 
where $\t_\star$ is the event horizon (\ref{rorstar}). 
Using (\ref{drel}), the latter can be rewritten as 
\be
d\eta\approx c_2(\t_\star) h(\t_\star)^{-1} d\d_1\,,
\label{nu2}
\ee
or 
\be
\eta(\t)\approx c_2(\t_\star) h(\t_\star)^{-1} \d_1(\t)\,.
\label{nu3}
\ee
Now,  rescaling the time 
\be
\tilde{t}=t\ { c_1(\t_\star) h(\t_\star)\over c_2(\t_\star)}\,,
\label{trescale}
\ee 
the metric becomes
\be
ds_{10}^2\approx \eta^2 d\tilde{t}^2-d\eta^2
-c_1(\t_\star)^2 d\bar{x}^2-9 c_2(\t_\star)^2 ds^2_{T^{1,1}}\,.
\label{resm}
\ee
From this form of the metric it is easy to see that if we Wick rotate 
$\tilde{t}$, we will avoid a conical singularity if we identify the 
Euclidean time  $i \tilde{t}$ with period $2\pi$. The periodicity 
in Euclidean time is identified as the inverse temperature. Tracing 
back to our original coordinate system, we identify the black 
hole temperature to be 
\be
T_H={c_1(\t_\star) h(\t_\star)\over 2\pi c_2(\t_\star)}\,.
\label{tH}
\ee
If $L\gg r_0$  (\ref{lr0}) and (\ref{rrs}) holds, 
we can estimate $T_H$ using asymptotics of the 
metric
\be
T_H\approx {r_\star\over \pi L^2}\ \left({r_0\over r_\star}\right)^4 
\ln^{-1/2}{r_\star\over r_s}\,,
\label{tHf}
\ee
where 
\be
r_\star\equiv \e e^{\t_\star/3}\,.
\label{rstar}
\ee
For the following estimates we take $r_\star\approx r_0$ (\ref{rorstar}).

Next, we compute the Bekenstein-Hawking entropy of the 
geometry (\ref{ktt}). We find the 8-dimensional area of the event 
horizon $\cA_8$ of the black hole to be
\be
\cA_8=c_1(\t_\star)^3 \left(3 c_2(\t_\star)\right)^5 V_3\ \omega_{T^{1,1}}
\approx {16\pi^3 \over 27} V_3\ r_0^3 L^2\ \ln^{1/2} {r_0\over r_s}\,,
\label{ahor}
\ee
where  $\omega_{T^{1,1}}=16 \pi^3/27$ is the area of the $T^{1,1}$ space 
(\ref{t11}), and $V_3$ is the 3-dimensional volume.
The entropy of the black hole is then
\be
S_{BH}={\cA_8\over 4 G_N}\approx {1\over 54 \pi^3 g_s^2 \alpha'^4}
V_3\ r_0^3 L^2\ \ln^{1/2} {r_0\over r_s}\,,
\label{entropy}
\ee
or, using (\ref{tHf}) and (\ref{lr0})
\be
S_{BH}\sim M^4 V_3 T_H^3 \ln^2(T_H)\,.
\label{fent}
\ee
From the ordinary statistical mechanics we know that the energy 
is $dE=T dS$ and the free energy is given by $F=E-T S$. Thus, for our 
black hole we estimate
\bea
&&E\sim {3\over 4}T S\,,\cr
\cr
&&F\sim -{1\over 4} T  S\,,
\label{ef}
\eea
at high temperatures.

The first computation of entropy of a finite temperature gauge
theory and its gravitational dual was reported in \cite{GKPe}.
The analysis of \cite{GKPe} was limited to a free $\cN=4$ $U(N)$
Yang-Mills theory, which gravitation dual is realized by a system of 
coincident near-extremal D3-branes. In the canonical ensemble, 
where temperature and volume are the independent quantities, the 
temperature of the YM theory should be set to the Hawking temperature 
in the supergravity. The Bekenstein-Hawking temperature of black 
3-branes then agrees up to a factor of $4/3$ with the free field 
computation in the YM theory. We would like to compare the 
BH entropy of our black hole (\ref{fent}) with the entropy of the 
finite temperature Klebanov-Strassler duality cascade. 
As in \cite{GKPe}, we will do only free field theory estimate.

The entropy computation of  KS duality cascade is simple. 
At temperature $T$, a typical energy of a weakly coupled 
bosonic mode is ${3\zeta(5/2)\over 2\zeta(3/2)} T\approx 0.770269 T$ 
while that of a fermionic mode is 
${3\zeta(5/2) (4-\sqrt{2})\over 2\zeta(3/2)(4-2\sqrt{2})} T
\approx  1.70007 T$. 
Klebanov-Strassler analysis \cite{ks0007} suggests that 
at energy $\mu\sim T$ the weakly coupled description 
of the duality cascade is in terms of $SU(N(T)+M)\times SU(N(T))$ 
gauge theory with two chiral superfields in $(\overline{N(T)+M}, N(T))$
representation and two chiral superfields in 
$(N(T)+M,\overline{N(T)})$ representation. $N(T)$ is given 
by (\ref{nmu}).  Altogether we have $N(T)^2+(N(T)+M)^2-2$ vectors (each one
contributes two bosonic degrees of freedom),  $4 N(T) (N(T)+M)$ complex scalars
(each one contributes two bosonic degrees of freedom) and 
$N(T)^2+(N(T)+M)^2-2+4 N(T) (N(T)+M)$ Weyl fermions (each one contributes 
two fermionic degrees of freedom). Recalling that in four dimensions 
a bosonic mode contributes $2\pi^2 V T^3\over 45$ to the entropy of the 
system (a fermionic mode contributes $7/8$ of the bosonic mode contribution),
we find 
\bea
S_{gauge}&=&\pi^2 V \left[N(T)^2+M N(T)+{1\over 6} M^2-{1\over 3}\right] T^3
\cr
\cr
&&\sim M^4 V T^3 \ln(T)^2\,,
\label{gaugeent}
\eea
where in the last line we suppressed numerical coefficients. 
Once we identify the temperature of the gauge theory 
with the Hawking temperature of the dual gravitational 
background, the  leading temperature dependence of 
(\ref{fent}) agrees with the leading temperature dependence 
of (\ref{gaugeent}) up to a numerical coefficient. 
This agreement provides a nontrivial check on the 
duality cascade proposed in \cite{ks0007}.  

In the previous section we argued that a black hole in  KT
background must be large enough to hide the naked singularity.  
Using (\ref{tH}), the criterion (\ref{rrs}) can be represented
as 
\be
T_H> T_s\equiv {r_s\over \pi L^2}\ e^{\xi/8} \sqrt{{8\over \xi}} 
\label{ts}
\ee 
Eq.~(\ref{ts}) suggests that the temperature of the 
Klebanov-Strassler gauge theory duality cascade  
should be larger than $T_s$ for  chiral symmetry to be restored.
It is tempting to speculate that $T_s$ is in fact the 
temperature of this phase transition. To settle this question, we would  have 
to compare the free energies of the black hole solution 
reported here with the free energy of the black hole in the 
Klebanov-Strassler geometry \cite{ks0007}, both having the 
same Hawking temperature.

\section{Discussion}

In this paper we found an exact black hole solution 
in the Klebanov-Tseytlin geometry \cite{KT}.  
We propose that this black hole realizes gravitation dual of the 
$\cN=1$ gauge theory duality cascade recently 
discussed by Klebanov and Strassler \cite{ks0007}
in the high temperature phase. More precisely, we expect 
the solution to be relevant at temperatures above chiral symmetry 
breaking temperature in the gauge theory.   
We showed that the entropy of  KS duality cascade at high 
temperatures is reproduced by the  Bekenstein-Hawking entropy 
of the gravitational dual up to a numerical factor. 
The entropy computation on the gauge theory side strongly
relies on the RG ``logarithmic running'' of the rank $N$ of the 
$SU(N+M)\times SU(N)$ gauge theory, which is the main prediction 
of KS duality cascade. The agreement in the entropy 
computation thus provides a check for KS duality cascade.

There are several interesting future directions \cite{wip}.
First of all, it will be extremely interesting to find the 
generalization of the above construction to the Klebanov-Strassler 
geometry of the deformed conifold \cite{ks0007}. This would allow us 
to study chiral symmetry breaking of the gauge theory 
within supergravity. KS geometry describes zero temperature 
confining vacuum of the gauge theory. As  the temperature increases, we expect 
a thermal phase transition into deconfinig phase. 
Given a  black hole solution in  KS geometry, we would be 
able to study this phase transition and its relation 
to the chiral symmetry breaking.

Finally, in this paper we presented yet another example supporting 
the general idea of \cite{g0002}. Here, a ``good'' naked singularity 
of \cite{KT} can be hidden behind the black hole horizon. Recently
Pando Zayas and Tseytlin discussed the supergravity solution 
of 3-branes on resolved conifold \cite{zt}. 
Much like in \cite{KT}, the solution has  a naked singularity 
in the IR. The singularity is of the repulson type. At this 
stage it is not known how this singularity is resolved, or,
whether it can be resolved at all. The construction
of a black hole solution in geometry \cite{zt} might shed 
some light on this problem.

\section{(Added) On the horizon singularity of the non-BPS 
background constructed in the previous sections.}
   
The metric ansatz for the non-extremal generalization of the KT 
background is given by eq.~(2.31) above, which for convenience 
we reproduce here
\be
ds^2_{10}=c_1(\t)^2\left[\d_1(\t)^2  dt^2-d\bar{x}^2\right]
-c_2(\t)^2\left[\d_2(\t)^{-2} d\t^2+9\ ds^2_{T^{1,1}}\right]\,.
\label{metric1}
\ee
The main assumption in searching for the non-BPS generalizations 
of the KT geometry used here is that the three-form flux 
on $T^{1,1}$ continuous to be self-dual even for the non-extremal
solution. This guarantees that the dilaton does not run. The type 
IIB equations of motion then  determine (2.46), (2.47)
\be
\d_2(\t)\left[\d_1(\t)\right]'={A\over c_1(\t)^4 c_2(\t)^4 }\equiv h(\t)
={81 A\over \e^4} e^{-4\tau/3}\,,
\label{drel1}
\ee
where $A$ is the non-extremality parameter. 
The warp  factor $\d_1(\tau)$ is given by 
\be
\d_1(\t)=\sqrt{f\left(h\left(\t\right)\right)}
\label{d1}
\ee 
where $f$ satisfies (2.53)
\be
\left[f(x)\right]'={3\over 2}\ 
{\left(30x+ 2\sqrt{135 x^2+400 f(x)}\right) f(x)
\over 9x^2-40 f(x)}\,.
\label{fppp}
\ee
with the boundary condition $f(0)=1$. All the other
functions specifying the  metric and the fluxes are determined once 
$f(x)$ is known. In Section 2.3, this equation  
has not been solved analytically.

Using (\ref{drel1}) the metric (\ref{metric1}) can be written as 
\be
ds^2_{10}=c_1(\t)^2\left[\d_1(\t)^2  dt^2-d\bar{x}^2\right]
-c_2(\t)^2\left[h(\t)^{-2} (d\d_1)^2+9\ ds^2_{T^{1,1}}\right]\,.
\label{rm}
\ee
with the event horizon  determined  from 
equation\footnote{For regular (nonsingular) horizons condition (\ref{ev}) 
is equivalent to $\d_2(\t_\star)=0$, which was used in Section 2.3.}
\be
\d_1(\tau_\star)=0
\label{ev}
\ee  
or, equivalently 
\be
f(x_\star)=0
\label{f0}
\ee
Numerically, the solution to (\ref{f0}) was found for 
{\it finite} $x_\star\approx 0.55647$. From the asymptotic 
analysis of Section 2.3 it follows that for sufficiently 
small $A$ (with arbitrary nonzero $x_\star$) the metric warp factors 
$c_i$ are regular at $x_\star$.  Then, the radial coordinate determined from 
\be
d\eta=c_2(\t) h(\t)^{-1} d\d_1
\label{eta}
\ee
is well-defined near $x_\star$.
We can see immediately from (\ref{rm}) that in this case 
the horizon is non-singular.  

Above arguments crucially depend on the statement that $x_\star$
is finite. This is actually incorrect \cite{new}. In 
\cite{new} equation (\ref{fppp}) has been solved {\it exactly}
\bea
x\equiv x(u)&=&{4 a\over 3 b}\ e^{-4 a u}\ \sinh 4b u  
\cr
\cr
f(x)\equiv f(x(u))&=&e^{-8 a u}
\label{sol}
\eea
with $3a^2=5 b^2$ and $a=243 A/4 >0$.
From (\ref{sol}) we see that $x_\star=0$, and so 
$\eta$ of (\ref{eta}) is ill defined near horizon 
as $h(\t_\star)=x_\star$. So, rather then cloaking 
the naked singularity of the KT geometry, the 
horizon of non-extremal solution presented here 
actually coincides with it \cite{new}. Given 
the uniqueness of the non-extremal solution  
for the self-dual ansatz of the three-form fluxes,
the latter must thus be relaxed for the gravitational 
background dual to the high temperature phase of the 
cascading gauge theory of \cite{ks0007}. 
The system of second order equations in the radial variable
describing non-extremal generalizations of the KT background 
whose solutions may have regular horizons is given in
\cite{new}.

Finally, from the exact solution (\ref{sol})
\bea
&&x\sim e^{-4 u a (1-\sqrt{3/5})}\cr
&&f\sim e^{-8 u a}\,,\qquad {\rm as}\qquad u\to\infty 
\label{qqq}
\eea
or $f(x)/x$ vanishes exponentially as $u\to \infty$. 
This exponential suppression is hard to capture in 
numerical analysis.

\section*{Acknowledgements}

I would  like to thank Oliver DeWolfe, 
Christopher Herzog, Gary  Horowitz,
Sunny Itzhaki, Igor Klebanov, 
Andrei Mikhailov, Aleksey Nudelman, 
Joe Polchinski,  Leopoldo  Pando Zayas  and  Arkady Tseytlin
for illuminating discussions. I am especially 
grateful to Joe Polchinski for reading the manuscript.
This work was supported in part by NSF grants 
PHY97-22022 and PHY99-07949.



\end{document}